%% file: Bottleneck.tex
\documentclass[twocolumn]{aastex63}

\submitjournal{PASP.  Also, https://arxiv.org/pdf/2004.03511.pdf}

\shorttitle{A Transient Utility}
\shortauthors{Kulkarni}

\graphicspath{{./}{figures/}}

\begin{document}

\title{Towards An Integrated Utility for Optical Transients}

\correspondingauthor{S.\ R.\ Kulkarni}
\email{srk@astro.caltech.edu}

\author[0000-0001-5390-8563]{S.\ R.\ Kulkarni}
\affiliation{Astronomy \&\ Astrophysics 249-17\\
California Institute of Technology\\
Pasadena, CA 91108, USA}


\begin{abstract} 
The ongoing optical time-domain astronomy surveys are routinely
reporting fifty  transient candidates per night. Here, I investigate
the demographics of astronomical transients and supernova classifications
reported to the Transient Name Server in the year 2019.  I find
that only a tenth of the transients were spectrally classified.
This severe ``bottleneck" problem should concern astronomers and
also funding agencies. The bottleneck will get worse by a factor
of 20 (or more) once LSST comes on line. We need to fundamentally
rethink the purpose of surveys for transients.  Here, after undertaking
a detailed investigation of this issue I offer some solutions.
Going forward, astronomers will employ two different methodologies:
(1) multi-band photometric method which is well suited to the study
of very large, many tens of thousands, samples of faint transients;
(2) spectral classifications of thousands of bright transients found
in shallow and nightly cadenced wide-field photometry surveys and
transients associated with galaxies in the local Universe.  The
latter program, in addition to unearthing new types of transients
and offering astronomers opportunities to undertake extensive follow
up of interesting transients, is needed to set the stage for the
former.  Specifically, I suggest a globally coordinated effort to
spectrally classify a complete sample of bright supernovae ($\lesssim
19.5\,$mag) and transients within the local Universe ($<200\,$Mpc).
The proposed program is within reach --  thanks to the on-going
wide-field surveys, the development of novel spectrographs tuned
for classification, great improvements in throughput of spectrographs
and the increasing availability of robotic telescopes.
 \end{abstract}




\section{Background}
 \label{sec:Background}

Time domain astronomy at optical bands is one of the oldest areas
of study in astronomy. At the beginning of the last century the
focus was the study of variable objects (e.g., RR Lyrae, Cepheid
variables) and novae. These studies led to revolutionary findings
of a much bigger Galaxy and a larger Universe than had been considered.

In 1936, using an 18-inch Schmidt-type telescope at Palomar, then
a novel type of telescope, Fritz Zwicky started a systematic program
to study supernovae (SNe).  Two years earlier Baade \&\ Zwicky had
already concluded that SNe marked the end states of stars. In that
spirit, going forward, I will use the term ``transient" exclusively
for events  which happen only once. This definition automatically
excludes eruptive variables such as dwarf novae and supernova
imposters.

By the end of the last century the study of SNe at optical wavelengths,
primarily motivated by the potential use of SNe as extragalactic
yardsticks, had become a major area of research in astronomy.  The latter
culminated with the discovery, through the light curves of SNe of
type Ia and the redshifts of the host galaxies, of the accelerated
expansion of the Universe.  The systematic study of cosmological
SNe is one of the principal goals of the upcoming Vera C.\ Rubin
Observatory\footnote{The principal survey that the Rubin Observatory
will carry out is the Legacy Survey of  Space and Time Survey or
LSST.} and the Wide-field Infrared Survey Telescope (WFIRST) space
mission.

\citet{z74} provides a very readable summary of supernova searches,
starting from the Palomar 18-inch program  and going through the
sixties (and including the Palomar 48-inch Schmidt-camera telescope
now renamed as the Oschin Telescope).  Early attempts at moving
away from the photographic plate and manually operated telescopes
are summarized by \citet{cmc75}. In my view, any serious student
of SN searches should read these two classic papers.  Many of the
ideas mentioned in these two papers are still valid today.

At the end of the last century, the Lick Observatory Supernova
Survey (LOSS) ushered in the era of dedicated remote and subsequently
robotic SN searches \citep{flt+01}.  LOSS targeted bright galaxies
in the local Universe and, at its  peak, was discovering about one
hundred SNe per year.  R.\ Quimby, for his thesis, used a 45-cm
telescope with the then large field-of-view of 3.4 deg$^2$
to search for SNe towards clusters of galaxies and found the first
two members of a new class of extragalactic transients  --
super-luminous supernovae \citep{q06}.  The All Sky Automated
Survey-Supernova survey (ASAS-SN; \citealt{spg+14}), initiated circa
2012, continued the tradition of looking for bright SNe but by
undertaking a systematic search of the entire night sky (``blind"
or ``un-targeted" searches). For these three projects, the discovery
rate was modest enough (few hundred per year) that members of the
observing team visually inspected candidate events and announced
highly reliable candidates.  The candidates were bright enough
(${\rm \lesssim18\,mag}$) that spectral classification of the
candidates could be readily undertaken by the global SN community.

Separately, a number of wide-field surveys with CCD mosaics began,
initially for specific purposes (e.g. searches for MACHOs, high
redshift Ia supernovae, asteroids).  One of the early general purpose
CCD survey, the Palomar QUEST survey, was undertaken with the Palomar
Oschin telescope \citep{dbm+08}. It ran from 2003--2008 and was
succeeded by the Palomar Transient Factory (PTF; \citealt{lkd+09}).
At about the same time, the Catalina Real-Time Sky Survey (CRTS;
\citealt{ddm+09}), and PanSTARRS-1 (PS1; \citealt{tsl+12}) came on
line. 

The SN rate shot up to thousands per year.  For these surveys Machine
Learning (ML) came in the nick of time \citep{brn+12,wss+15}, as
it lead to a substantial reduction in the load of visual inspection
of candidates.  These surveys marked the beginning of the era of
``industrial" time domain surveys.  At the present time, the major
industrial surveys are Asteroid Terrestrial Impact Last Alert System
(ATLAS; \citealt{tdh+18}), PS1 and the Zwicky Transient Facility
(ZTF; \citealt{bkg+19}; the successor to PTF).

A major advance in the TDA field was the the development and immediate
implementation of a novel imaging differencing algorithm (aka ``ZOGY;
\citealt{zog16}).  In this method, the recent image and the reference
image are convolved with each other's PSF and then subtracted. The
resulting subtracted image, as one conclude based on the symmetry,
will have far fewer imperfect subtractions.This method is not only
optimal but also reduces the reliance on ML.  Every clear night ZTF
pumps out, in real time, hundreds of thousands of ``alerts"\footnote{In
the parlance of modern time domain astronomy, an alert is a
$n$-$\sigma$ change in RA, DEC or flux.  Taking the lead from LSST,
the ZTF team has elected to use $n=5$.} \citep{pbr+19}.

The industrial surveys with their tremendous output of transient
candidates have changed the landscape of the optical TDA field. The
purpose of this paper is to quantitively review this new landscape
and identify new opportunities, especially bearing in mind the arrival
of the Rubin Observatory.  

The paper is organized as follows. The huge growth in the transient
detection rate has led to a new clearing house for alerts which is
summarized in \S\ref{sec:Dissemination}.  The increase in the
transient detection rate has not been accompanied by corresponding
increase in spectral classification. In fact, most of the transient
candidates ``die on the floor of the dome".  Stated bluntly, transient object
astronomy field has already undergone a profound shift -- from a
relatively event-poor field to an event-drowning field. In view of
this fundamental change in \S\ref{sec:Discovery} I try to understand
what constitutes meaningful discovery in the field of optical transients.
In \S\ref{sec:2019} I present a quantitative analysis of ``Astronomical
Transients" (ATs) and SNe reported to the Transient Name Server
(TNS)\footnote{\url{https://wis-tns.weizmann.ac.il/}} in the year
2019. The bottleneck factor -- the ratio of reported ATs to spectrally
classified transients -- now stands at 10. This factor is expected
to increase over time.  This bottleneck not withstanding, I revisit
the fundamental importance of spectral classification and determination
of redshifts of transient events (\S\ref{sec:SpectralClassification}).
In \S\ref{sec:TransientUtility} I suggest an  integrated ``Optical
Transient Utility" consisting of an imaging element and a spectral
classification element.  The two elements together provide, routinely,
nightly cadenced light curves of the sky accompanied by spectral
classification of all bright transients.  In  \S\ref{sec:Conclusions},
I summarize the principal conclusions and then follow it up with
some observations on the importance of clearing houses and
astroinformatics.

\section{Dissemination in the era of industrial surveys}
 \label{sec:Dissemination}

Even a decade ago it was customary to announce SN discoveries via
the Central Bureau for Astronomical Notifications
(CBAT).\footnote{\url{http://www.cbat.eps.harvard.edu}}   Astronomers
sent reports to an editor and the editor assigned a SN name (e.g.,
SN\,1987A). The standard medium for dissemination was telegrams of
yore, later replaced by elegantly laid out post cards  and most
recently expressed through ``electronic telegrams".

Noting the rapid changes in the SN field in 2009 (\S\ref{sec:Background})
\citet{gmm+13} reviewed the operational side (detection rate, modes
of reporting, classifications) of the SN field for the year 2010
and 2011.  In 2009 the number of classifications was slightly behind
the number of candidates. However, they found that over the next
two years the discrepancy increased.  Separately, the number of SNe
(by which I mean transients with secure spectral classification)
started to rise: 538 SNe in 2010 to 926 SNe in 2011. The reporting
mechanism became chaotic with some observers reporting to the CBAT,
some to Astronomer's
Telegram\footnote{\url{http://www.astronomerstelegram.org/}} (ATEL)
and others via their own project pages.

The explosive growth in the detection rate of transients  and the
chaotic reporting led the International Astronomical Union (IAU)
to call upon the community to create a single entity to receive
reports of transients and assign them official IAU names.  The
result was the Transient Name Server (TNS, introduced in the previous
section) which is run by a group of public spirited astronomers at
the Weizmann Institute of Science, Israel. The transition from CBAT
to TNS took place on 1 January 2016.

It is now the standard practice that astronomers routinely report
(hopefully high fidelity) transient candidates to TNS and in return
get  ``Astronomical Transient" (AT) designations for their candidates.
The ATs are indexed in the same way as SN (e.g., AT\,2019A,
AT\,2019B,..., AT\,2019Z, AT\,2019aa, AT\,2019ab,...).

\section{The Discovery Process}
\label{sec:Discovery}

We use the word ``discovery" routinely. However, it is a potent
word and it is worth dissecting the intended meaning of this word.
In this section, I first discuss the field of GRBs as a case study
and then follow it up summarizing the current discovery process for
supernovae.

\subsection{GRBs: A Case Study}
 \label{sec:GRBs}
 
GRBs have distinct high energy temporal and spectral signatures.
It is relatively easy to reject false positives such as flares from
the Sun or gamma-rays from lightning.  In fact, the gamma-ray
spectral and temporal data are sufficient, in most cases, to classify
sub-types (short-hard, long-soft, repeater, tidal disruption events)
within the family of gamma-ray transients.  Next, the total annual
rate of GRBs is modest, no more than a thousand per year.

The first detections of flashes of gamma-rays were made by Vela-3
and Vela-4 satellites in 1967. However, evidence that these are not
artifacts nor of local origin (Earth, Sun) came only with triangulation
offered by a constellation of Vela satellites (5A, 5B, 6A and 6B)
circling the earth \citep{kso73}. It is at this point one could say
that extra-solar gamma-ray bursts were discovered.  Over the next
thirty years progress was stymied by poor localization and so the
main focus was on statistical studies (e.g., log$n$-log$s$ and
distribution on the sky). The discovery of long-lived lower-energy
emission radiation, aka the ``afterglow" \citep{cfh+97}, revived
the field.  Thus, after 1997, what mattered the most was whether
an afterglow was detected or not. In effect, the limiting step in
GRB astronomy was set by the detection rate of afterglow emission.

The {\it Swift} Observatory\footnote{\url{https://swift.gsfc.nasa.gov/}.
Now renamed as the Neil Gehrels Swift Observatory.} was explicitly
built to detect X-ray and UV/optical afterglow.  To this end, it
employed a space-craft with high maneuverability. A wide-field
coded-aperture mask imager detected  bursts of hard X-rays with
``crude" localization (arc minutes), following which the space-craft
rapidly slewed so that ``narrow" field-of-view but high sensitivity X-ray
and optical/UV imagers could start detecting the afterglow.  Armed
with  arc-second localization of the afterglow astronomers rush off
to undertake multi-wavelength and, in particular, obtain the redshift
via the afterglow (ideally) or that of the host galaxy (at a later
time). It is this triplet of efforts that has enabled {\it Swift}
to advance the GRB field.

In fact,  a future general-purpose GRB missions will have to at least
match {\it Swift}'s afterglow capabilities in order to get funded.
Indeed, the planned Chinese-French Space Variable Objects Monitor
(SVOM)\footnote{\url{http://www.svom.fr}} carries wide-field GRB
detectors, narrow-field but sensitive X-ray and optical telescopes
and, in addition, an ambitious array of dedicated ground-based
optical telescopes (one for wide field imaging and 1-m telescopes
with three-band imagers).

The essential point here is simple: the value of a discovery is
relative.  In the nascent phase of a subject, initial secure
detection(s) are trail blazing.  As the subject progresses, mere
detections while also termed as discoveries constitute increasingly
poor return of understanding. In order to make advances, additional
data (e.g., afterglow) are needed to make the detected candidates
to have some value.  In that sense the process of ``discovery"
becomes longer (and usually distributed over several efforts).

I conclude this section with two points. First, there is a striking
parallel between the historical development of the GRB field and
that of Fast Radio Bursts (FRBs), except the development of FRB
field is proceeding a much faster rate, relative to that of GRBs
\citep{k18}. Second, I note that over the course of the mission
lifetime, {\it Swift} now essentially acts like ``GRB Utility\footnote{As
in a  water or electric utility.}" -- a reliable source of usable
products (time of explosion, arcsecond position and early afterglow
light curves).  In a fundamental way, this transformation is a
direct indicator of the maturity of the field.  After having studied
over several thousand GRBs, even an astronomer who is devoted to
GRBs would not be willing to chase every GRB. Instead, astronomers
review the Swift data products and define sub-samples that are
worthy of further study.  I predict that in less than five years
the field of FRBs will have their own ``FRB Utilities".

The continuing success of the aging {\it Swift} mission motivates
me to suggest here an ``Optical Transient Utility" -- a reliable
supplier of nightly cadenced light curves along with spectral
classification for all transients satisfying a brightness flux limit
or a distance limit. Astronomers with patience will use the large
data base to quantitatively model the transient phenomena (rates,
subtypes, yields and so on) whereas interested in action will be
on the lookout for unusual events or selected samples to undertake
deeper (follow up) studies.  This Utility idea is fully developed
in \S\ref{sec:TransientUtility}.

\subsection{The Discovery Process for Supernovae}
 \label{sec:DiscoveryProcess}

For extragalactic optical transients, unlike GRBs, it not possible
to make firm identifications of the nature of the transient based
on a few initial multi-band photometric points.  Considerable
additional information, either in the form of a classification
spectrum or an extensive multi-band light curve is needed to securely
classify the transient. In some sense, this is the equivalent of
the information supplied by the afterglow phenomenon. Finally, to
set the energy scale of the event, we need to know the redshift (or
distance) to the event.

Next, the transient optical sky is more fecund than the transient
gamma-ray sky.  For instance, the annual all-sky rate for supernovae
is about 32000 for peak magnitude of 20 \citep{fnr+19}. Going even
a magnitude fainter will dramatically increase this rate. These
rates are so large that it is a capital mistake to measure the
anticipated impact of a new facility by the expected yearly haul
of transient candidates.

In my view, going forward, a {\it useful} discovery of a SN involves
four distinct steps which are discussed below.

\subsubsection{Detection of an event}
 \label{sec:Detection}
 
The first step is {detection of an event. At the present time, this
step is provided by imaging surveys.\footnote{Though, as time goes
by, DESI and other massively multiplexed spectroscopic surveys,
will give rise to a cottage industry of ``spectral transient
candidates".} The latest image is compared to a reference image and
sophisticated algorithms employed to reject imaging and instrumental
artifacts and poor subtractions. The next step is to determine
whether the event is a transient.

\subsubsection{Purity of the Candidate} 
 \label{sec:Purity}

Before addressing the technical details  I want to clearly address
an important but sticky issue: ``what is a transient?". All astronomers
would agree that  asteroids or variable stars are not transients.
All astronomers would also agree that the term transient should be
applied to singular (``one-off") events which are accompanied by a
dramatic rise in the flux.  From decades of study we know that such
events involve wholesale destruction of an object or a massive
rearrangement of the system. By this criteria, at the present time
the set of transients include supernovae of all sorts, stellar
mergers of all sorts and tidal disruption events (TDEs).

However, there is likely to be some dispute about eruptive variables:
classical novae and related variables (e.g., dwarf novae), type I
bursts, soft x-ray transients, x-ray novae, supernova imposters and
flare stars (and related phenomenon) variables. In these objects
the events arise from surface explosions, rapid changes in accretion
rate or magnetic reconnection.  I advocate that we call such violent
but recurrent events as eruptive variables. 

The second step is to determine the confidence in the detection of
the transient -- the purity.  The false positives are moving objects,
variable stars and eruptive stars.  To this end, the observing team
would likely have employed one or more strategies to cull out moving
objects (e.g., by undertaking two observations, separated by an
hour).\footnote{At the bright end, say 18\,mag, the asteroid catalog
is quite complete and so one could review the candidate against the
ephemerides of asteroids.} Quiescent counterparts are expected for
variable stars including eruptive variables.  Deep multi-band images
(e.g., PS-1, WISE, GALEX) can be used to search for the counterparts
and thus filter out the eruptive variables.  Finally, contextual
information (e.g., a diffuse object next to a transient is likely
to be the host galaxy) can be used to enrich the transient yield.

The interested reader is directed to Appendix~\ref{sec:CriteriaATs}
where I provide summaries of the classification strategies employed
by the major groups involved in ongoing optical TDA surveys.

\subsubsection{Announcing the candidate}
 \label{sec:Announcing}
 
Some groups are internally organized to both detect, classify and
follow up the sources. Examples include dedicated type Ia supernovae
and private surveys carried out with ZTF. For groups or surveys
without such built-in capability, prompt communication of the
transient opens up the possibility of follow up (multi-band,
multi-``messenger") astronomy. At the present time, this step is
accomplished by astronomers uploading the vital statistics of the
transient to TNS.

In this context I bring up the possibility that there could be two
different epochs: the  epoch of the ``first detection"  and the
epoch at which the event is reported to the TNS. Some groups may
be adventurous and report the first detection whereas others may
wish to see a second detection before uploading a candidate to TNS.
Given this situation, it would be helpful for the community to have
both these events included in the AT data base. However, the time
honored and the most fair way to settle issues of ``claim" is to
link the AT to the group which reported first.   I will admit that
this rule is likely to tempt some astronomers to register events
with modest significance.  A solution  is for a ``ratings" group
to keep a track of the quality of the events reported by various
groups or brokers (and addressed in \S\ref{sec:ClearingHouses}).

\subsubsection{Linking the transient to known classes}
 \label{sec:Linking}
 
The last step is to decisively link the transient to a major class
of explosions.  In almost all cases,  a quality spectrum leads to
definitive classification. SN spectra are sufficiently unique to
distinguish them from flaring stars and Galactic eruptive variables.
Separately, spectra provide a decent estimate of the redshift which
sets the physical scale of the event. Other alternatives include
multi-band light curves, which are discussed in
\S\ref{sec:SpectralClassification}.

Taken together, the four steps described above constitute the
``discovery" of an extragalactic optical transient. Given this
multi-stage effort, it is important to credit ``discovery" in a
fair and equitable way. It will be increasingly the case that the
detection of the transients will be undertaken and disseminated  by
one or more surveys, the alerts will be processed for purity by one
more ``brokers" (or their subscribers) and finally one more observers
will undertake key follow up observations.

\section{Astronomical Transients \&\ Supernovae in 2019}
 \label{sec:2019}

From the TNS website, using a program I had written \citep{k20}, I
downloaded, on 22 March 2020, the AT master catalog for  the year
2019.  I found a total of 18296 entries of which  2012 were
classified as SNe. Each AT is summarized by a record which consists
of 24 fields.  At this point in time, all candidate events are found
via imaging (photometric) surveys.  Some surveys operate without
filters whereas others use traditional filters. It is reasonable
to approximate all reported magnitudes as ``V" band.


\begin{deluxetable}{lr}
\tablecaption{AT candidates sorted by survey telescopes}
\tablewidth{0pt}
\tablehead{
\colhead{Imaging Telescope} & \colhead{$n$}
}
\startdata
\input{AT_statistics.txt}
\enddata
\tablecomments{$n$ (column 2) is the number of ATs first reported
by telescope (column 1).  The histogram has been cutoff at number
of ATs,  $n<100$. }
 \label{tab:AT2019}
\end{deluxetable}

The first four fields are basic: ID (record index), AT Name, RA and
Dec.  A number of key words are related to photometric detection:
the name(s) of photometric surveys (``Discovery Data Source/s",
field 10), the internal name of the source (``Discovery Internal
Name", field 13), the name of the imager (``Discovery Instrument/s",
field 14), the magnitude upon the first detection (``Discovery
Mag/Flux", field 19), the ``Discovery Filter" (field 20) and the
date of the first detection (``Discovery Date", field 21).  The
``Reporting Group/s" can be found in field 9.

The fields related to spectroscopic observations are as follows:
field 6 (``Redshift" of the transient), field 8 (``Host Redshift"),
field 11 (``Classifying Group/s") and field 15 (``Classifying
Instrument/s"). In almost all cases a spectrum leads to a secure
classification of the event and is recorded in field 5 
( ``Object type").


\begin{figure}[ht!]
 \plotone{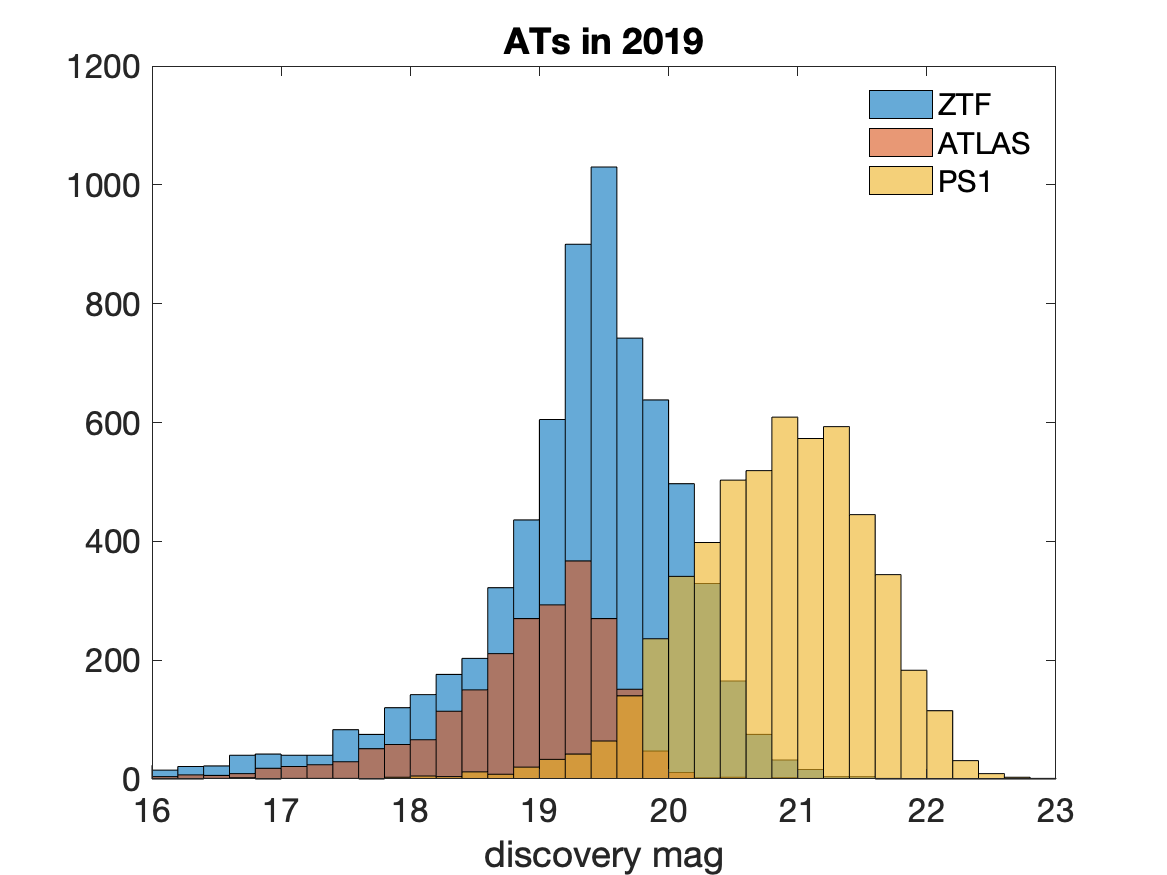}
  \caption{Histogram of the magnitudes upon detection of ATs reported
in 2019 for ZTF, ATLAS and PS1. ATLAS and PS1 self report
(\S\ref{sec:PS1ATLAS}).  Several groups receive ZTF alerts and
submit their own reports: Bright Transient Survey (\S\ref{sec:ZTF-BTS}),
AMPEL (\S\ref{sec:AMPEL}) and ALeRCE  (\S\ref{sec:ALeRCE}).  }
 \label{fig:AT2019a}
\end{figure}


\begin{figure}[ht!]
 \plotone{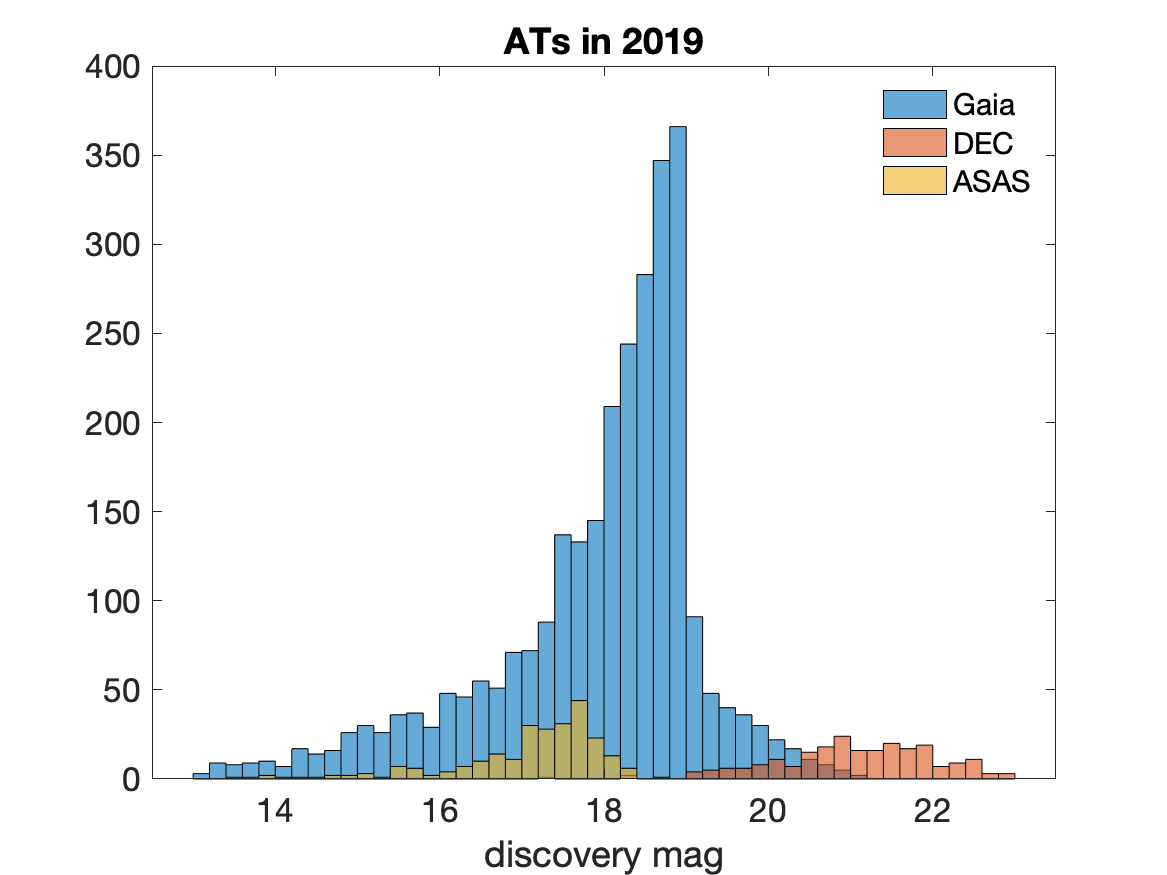}
\caption{The same as in Figure~\ref{fig:AT2019a} (but with a reduced
y-axis scale) for {\it Gaia} (\S\ref{sec:Gaia}), DECam (in the
course of pursuing GW events) and ASAS-SN.  }
 \label{fig:AT2019b}
\end{figure}

An inspection of the 2019 data shows the following distribution for
Object Type: null value (15987 events), SN (1932 events), TDE (5
events), ILRT (Intermediate Luminosity Red Transients; 1 event) and
a total of 101 events covering the set [AGN, CV, Galaxy, LBV,
Light-Echo, M dwarf, Nova, QSO, Varstar].  Only for events with
Object Type=SN will the forename change from AT to SN.  No equivalent
rechristening takes place for any other transients such as TDE or
ILRT.\footnote{``As of this point" -- Ofer Yaron, Curator for TNS.}

Using the first name\footnote{The field can contain several ``discovery
data" sources, separated by commas.}  in field 10 I link each AT
to its discovery survey.  The resulting histogram of candidates
with respect to the originating surveys can be found in
Table~\ref{tab:AT2019}.  As can be seen from this Table, the dominant
contributors are ZTF, PS1, {\it Gaia} and ATLAS.
The histogram of the magnitude of the first detection is shown in
Figure~\ref{fig:AT2019a} and Figure~\ref{fig:AT2019b}.   From this
Figure, it is clear that PS1 is the most sensitive survey followed
by ZTF, ATLAS and {\it Gaia}.

One of the unexpected surprises is that supernovae can inform us
of how complete are our catalogs of galaxies. The luminosities of
galaxies can vary over eight orders of magnitude whereas supernovae,
especially type Ia, are famous for their constancy (and brilliance).
Thus, supernovae are excellent beacons of galaxies and with a modest
effort, type Ia supernovae can be used to make precise assessment
of the completeness of catalogs of nearby galaxies \citep{kpm18}.
In view of this connection, it is interesting to report plausible
host galaxies of supernovae. In fact, this information can be found
in field 7 (``Host Name"). Analysis of the host names is presented
in \S\ref{sec:HostNames}.

\subsection{Statistics of Classified Events (Supernovae)}

An inspection of the TNS master catalog for 2019 shows that only
2012 of the ATs were classified as SNe. The histogram of the discovery
magnitudes of the SNe is shown in Figure~\ref{fig:SN2019}.

In Table~\ref{tab:ClassifyingSpectrographs} I provide a histogram
of the telescopes/instruments which undertook these classifications.
The distribution in this table is not necessarily reflective of the
sensitivity of the spectrographs but also of the priority accorded
to spectral classification. For instance, the EFOSC2+NTT has
considerable capacity to classify but the key project
ePESSTO\footnote{\url{http://www.pessto.org}}, which has been granted
90 nights per year,  has also in its ambit detailed spectral studies
of a variety of transients. SPRAT on the Liverpool Telescope and SNIFS
on UH88 can become major players if more
telescope time can be allocated to spectral classification.


\begin{figure}[ht!] 
 \plotone{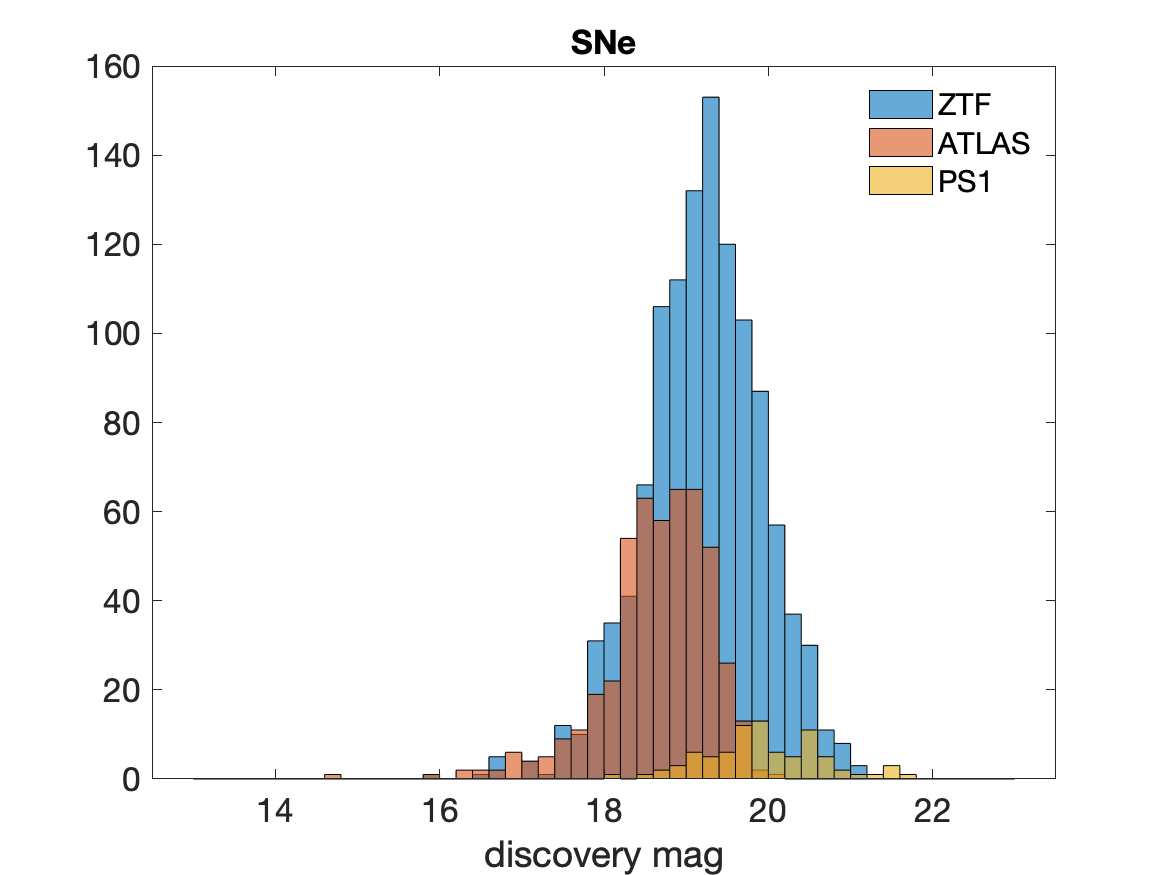} \caption{As in
Figure~\ref{fig:AT2019a} a histogram of the magnitudes upon
discovery but of classified SNe.  }
  \label{fig:SN2019}
\end{figure}

The SEDM (Spectral Energy Distribution Machine or SEDM) was designed
to solely classify supernovae. Extragalactic transients have
significant expansion velocities ($v/c \gtrsim 0.01$) and so
a spectral resolution, $R=\lambda/\Delta\lambda$ of 100 is sufficient
for classification.  Thus the SEDM with $R\approx 100$ trades 
spectral resolution for telescope aperture (see  \S\ref{sec:SEDM}
for details). It is for this reason that the SEDM mounted on a mere
(but dedicated) 60-inch telescope is able to account for 40\% of
the global SN spectral classification (see \citealt{fms+19} for the
target throughput of SEDM).


\begin{deluxetable}{lrr}
\tablecaption{Classifying Instruments \&\ Telescopes}
\tablewidth{0pt}
\tablehead{
\colhead{Telescope-Spectrograph} & \colhead{$n_s$} &\colhead{\%}
}
\startdata 
 \input{TelescopeSpectrograph.txt} 
\enddata 
 \tablecomments{$n_s$
(column 2) is the number of spectra observed with the
telescope-spectrograph (column 1) whereas column 3 is the ratio of
$n_s$ to the total number of classification and expressed as a
percentage.  Here, ``anon" refers to those spectra for which there
is no telescope-spectrograph entry. The histogram has been cutoff
at $n_s<10$. }
 \label{tab:ClassifyingSpectrographs}
\end{deluxetable}

\subsection{The Bottleneck}

My analysis shows 1800 events have only one classification spectrum,
202 have two spectra and nine have three spectra. SN\,2019clp is
distinctive for having six spectra.  Figure~\ref{fig:AT_SN2019}
graphically summarizes the ``bottleneck" problem.  The  bottleneck
factor, defined as the ratio of the identification of transients
to spectral classification is nearly 10.


\begin{figure}[ht!] 
 \plotone{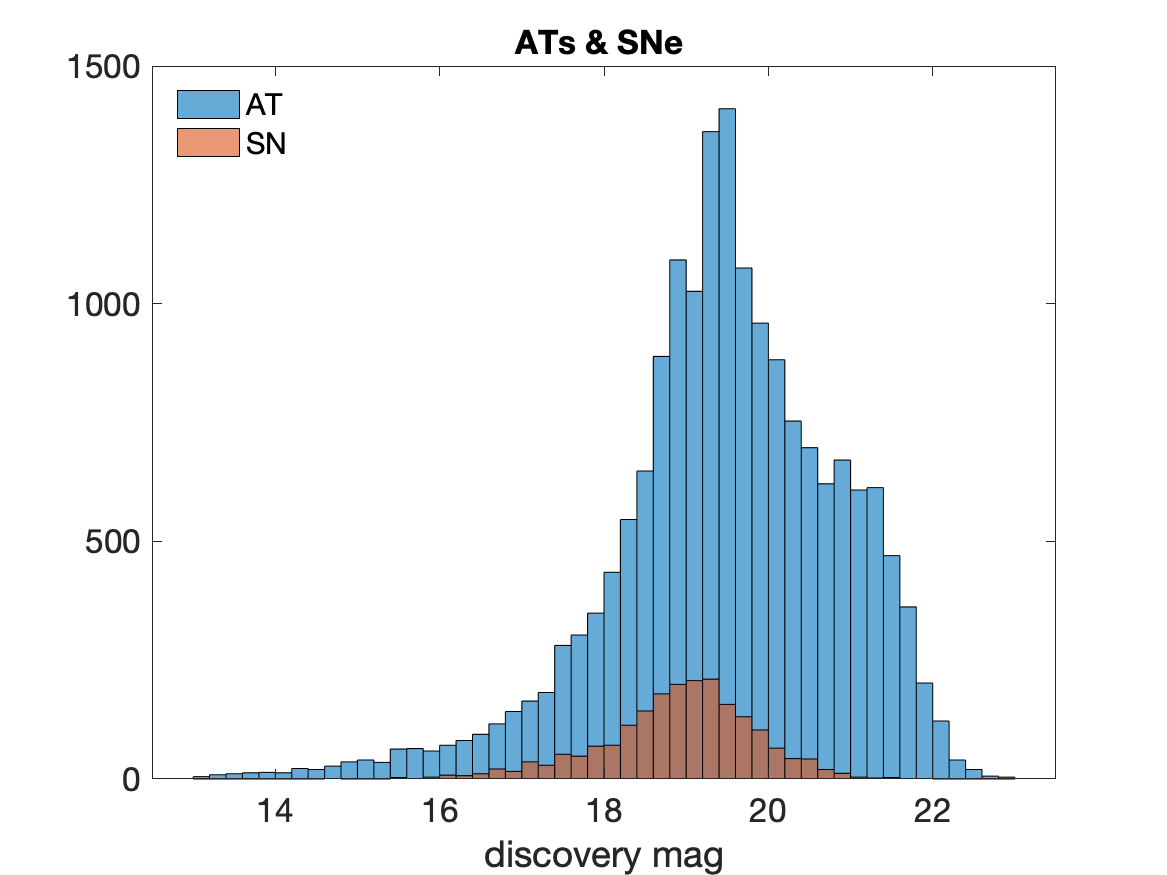} 
 \caption{As in
Figure~\ref{fig:AT2019a} a histogram of the magnitudes upon
discovery of all ATs and SNe reported in 2019. }
  \label{fig:AT_SN2019}
\end{figure}

An astute reader is likely to be surprised by the divergence between
ATs and SNe at the bright end.  I present an analysis of the origin
of this divergence in \S\ref{sec:BrightTransientCandidates}.

\section{Spectral classification \& Redshift: Foundational Values}
 \label{sec:SpectralClassification}

Going forward, we assume that improvements in image differencing
algorithms and increasingly sophisticated filtering algorithms would
have gotten rid of events arising from instrumental artifacts and
that archival and contextual data will be used, along with sophisticated
Machine Learning algorithms, to better classify events: asteroids,
Galactic variables, flare stars, eruptive variables, burbling AGN
and extragalactic explosive events.

\begin{deluxetable}{lrr}[hbt]
\tablecaption{Local Volumetric Annual Rates of Extra-galactic Explosions}
\tablewidth{0pt}
\tablehead{
\colhead{Class} & \colhead{$\mathcal{R}$}&\colhead{Ref.}\cr
\colhead{}          & \colhead{$({\rm Gpc^{-3}\,yr^{-1}})$}& 
}
\startdata
Thermonuclear SNe & $3\times 10^4$ &  [1] \\
Core-collapse SNe & $7\times 10^4$   &  [2] \\
SLSN &  $\approx 100$&[3]\\
TDE  &    $\approx 700$& [4]\\
LLGRB & $\approx 200$ & [5]\\
LGRB$^\dagger$ & $\approx 3$  &[5] \\
SHB$^\dagger$ & $\approx 2$ & [5,6]\\
BNS & $<800$ & [7] \\
\enddata
 \tablecomments{The  annual volumetric rates at
$z\approx 0$. Thermonuclear supernovae cover the entire class of
type Ia supernovae and are expected to be powered by fusion.
Core-collapse events can have hydrogen but their power source is
ultimately gravitational collapse.  Some of them could be powered
by subsequent engine activity. The acronyms are as follows:
super-luminous supernovae (SLSN), tidal disruption events (TDE),
low luminosity gamma-ray bursts (LLGRB), long duration gamma-ray
bursts (LGRB), short hard bursts (SHB) and binary neutron star
coalescence (BNS).  $^\dagger$No correction for beaming. The
multiplier could be as high as 100. \hfil\break
 {\it References:}
[1] \citet{llc+11}.  [2] \citet{hpk20}. The total rate is from
\citet{llc+11} but this reference gives a break down between II,
Ibc, Ic-BL etc.  [3] \citet{qya+13}.  [4] S.\ van Velzen (pers.
comm.).  [5]  \citet{szl15} [6] \citet{dto+20}.  [7] \citet{kns+17}.
}
  \label{tab:Rates}
\end{deluxetable}

From Table~\ref{tab:Rates} we see that SNe dominate the demographics
of extragalactic explosions.    So the path to all other transients,
especially the rare transients, requires a masterful understanding
of SNe, if only to eliminate them in the quest for, say, kilonovae.

\citet{m41} introduced classification  based on spectroscopy. The
initial classification was type I (Hydrogen absent) and type II
(Hydrogen present).  Over time, spectral sub-types (Ib,Ic, IIn) and
light curve based subtypes (IIL, IIP) were introduced (see
\citealt{f97}).  For completeness, at the suggestion of my colleague,
Anders Nyholm, I mention the pioneering work of Piotr Kulikovsky
(1910--2003) who published a classification system also based on
light curves (\citealt{k44};  see \citealt{ln83} for a modern
citation). This system distinguished between four types of supernovae
and found that some classes of supernovae favored star-forming
galaxies and others late-type galaxies.

I offer the following {\it ansatz}: about one hundred data points
are needed to make a fairly secure classification. These points
could either be a low resolution optical spectrum or a multi-band
light curve with tens of epochs properly sampled.\footnote{For
instance, for a flare star, you would need sampling on timescales
of minutes to hours and for SLSN, sampling on timescales of weeks
to months is adequate.} However, in either case we need to know the
redshift of the transient. A spectrum of the transient will not
only provide a robust classification\footnote{For obdurate sources,
a repeat spectrum usually clarifies.} but also readily result in
the redshift of the transient (or the host).  For high redshift
transients, say $z\gtrsim 0.3$, multi-band photometry of the host
galaxy will be adequate to get a useful estimate of the redshift
\citep{cl04} and thus set the energy scale.

The above discussion suggests a bifurcation in the methodology of
the study of transients: the photometric time series method which
is well suited to the numerous but faint transients,\footnote{In
fact, the Transients and Variable Stars science working group of
LSST has initiated a major program to precisely address this issue:
Photometric LSST Astronomical Time-Series Classification Challenge
(PLAsTICC); see \url{https://plasticc.org}.} and spectral classification
for bright transients, which will be necessarily fewer relative to
the population of faint transients.  The faint transients will be
largely at high redshifts (for which the photometric method will
provide a usable redshift) while the bright transients, in each
category, will  be at low redshifts.

\section{A Public Transient Utility}
 \label{sec:TransientUtility}

Earlier we noted that there has been a profound shift in transient
object astronomy, namely, astronomers are already seemingly overwhelmed
by the transient candidate discovery rate. Next, we found that
spectral classification observations for only a tenth of the transient
candidates are being undertaken.  I review the near-term landscape
whilst keeping these two conclusions in mind.

The next major development in the TDA field is LSST.\footnote{Given
the delays due to COVID-19, a plausible schedule for full operations,
including routine real-time release of alerts, of the Rubin Observatory
is early 2024.}  The great strengths of LSST are (1) photometric
accuracy and precision, (2) superior astrometric precision for faint
objects and (3) accurate knowledge of the point spread function.
As a flagship survey, LSST has to respond to many constituencies.
The resulting slow cadence of LSST is well suited for the study of
slowly varying transients (such as SLSN) including cosmological Ia
supernovae, lensed supernovae, thanks to the time dilation factor
of $1+z$.  There are plans to set aside a small amount of time for
moderate to high cadence studies of selected fields. Given the high
density of faint events, such ``mini" surveys are well suited to
explore the phase space of  faint and fast events. In particular,
with these mini-surveys, LSST, because of its depth, is well
positioned to undertake a comprehensive investigation of relativistic
stellar explosions and related phenomenon (especially orphan
afterglows).

In reverse, the areal density of objects decreases with increasing
brightness. Thus, for the brightest phenomenon one needs an all-sky
survey. At the present time, this is the value that ASAS-SN provides,
given the on-going wide-field surveys (ATLAS, PS-1, ZTF).  In the
same way, in the LSST era, we need to have wide-field nightly
cadenced  sky survey(s) to operate in conjunction with LSST. Accepting
this conclusion and motivated by the success of  {\it Swift} ``GRB
Utility" (\S\ref{sec:GRBs})  I propose an integrated ``Optical
Transient Utility".  This  facility will, night after night, produce
densely sampled light curves along with the ``ground-truth" provided
by spectral classification.  The resulting data will be complementary
to LSST and, equally importantly, produce the foundational data for
Machine Learning algorithms which can then be applied to large
samples of faint transients with modest number of photometric points.
Separately, other astronomers could undertake detailed studies of
interesting objects (based on light curves or spectral classification
or association with other surveys such as the Spektr
RG\footnote{\url{http://www.iki.rssi.ru/eng/srg.htm}}).

As with {\it Swift} all data will be released in near-real time.
For rapidly evolving transients (e.g., young supernovae) it is
critical that the alerts carry a light curve history with force
photometry (including rigorously specifying upper limits). Spectral
classification observations will be undertaken for a well-defined
sample of candidates and  data released in real time.    A
monthly release of archive quality images and light curves would
be helpful for stellar astronomy.  

A facility as described above, especially with the listed
demands, will come with a cost.  However, the cost can
be justified by noting that the utility will serve the entire
astronomical community.  Below, I discuss two specific key projects
which form the backbone of the proposed Transient Utility: a
flux-limited sample and a volume-limited sample.\\

\noindent{\bf Bright Transient Survey (BTS).} Flux-limited surveys
have been a corner stone of astronomy.  A recent example of such a
survey is the ZTF BTS \citep{fms+19}.  The estimated annual all-sky
supernova rate (Ia and core-collapse) is [2200, 4300, 8400, 16231,
31478] for peak magnitude of [18, 18.5, 19, 19.5, 20] \citep{fnr+19}.
A large sample would allow for detailed studies relating supernova
types to host galaxy parameters. A large sample would result in a
bigger ``ground truth" data base for ML algorithms.  Finally, the
large size of the sample permits intense searches for rare types.
A sample focused on nearby transients allows for the discovery of
watershed events such as AT2019cow \citep{pgs+18}.

Supernovae of type Ia have played an outsized role in the development
of modern astronomy. The determination of large scale mass distribution
is a major goal of modern astronomy. In a purely Hubble flow, the
velocity of the host galaxies would be the same as that computed
from the Hubble law applied to the distance given by Ia SNe.  However,
large fluctuations in matter (dark and otherwise) result in the two
velocities being discrepant with the difference velocities (“peculiar
velocities”) acting as tracers of matter.  As noted by many authors
this opens up the possibility of constructing the local matter
makeup (e.g., \citealt{s87,gls07,hdd+17}).  In fact, in this regard,
there are several on-going efforts but based on using the Fundamental
Plane relation for elliptical galaxies (6dFGS, TAIPAN) or Tully-Fisher
relation for spiral galaxies (e.g., WALLABY on ASKAP).

The strength of SN Ia sample is the precision of the distance
estimate and the weakness (relative to the galaxy methods) is the
smaller sample size. Several recent studies have investigated this
method in considerable detail and find that the SN approach is
promising provided the SN sample is in the many thousands (see, in
particular, \citealt{aof19,grr+20,kl20}).  In addition to determining
the mass make up on local scales (comparable to the Baryon Acoustic
Oscillation, BAO, scale), the same observations can determine the
growth index, $\gamma$ where $f\propto \Omega_M^\gamma$ with $f$
is the linear growth rate and $\Omega_M$ is the ratio of the matter
density to critical density.

The main issue is the size of the sample. A sample size of 6000
type Ia supernovae will more than match the  planned galaxy surveys
(Jakob Nordin, pers.\ comm.).  ZTF alone is already classifying
nearly a thousand Ia supernovae per year.  If we stick to 19.5 mag
and only consider supernovae which go off in the night sky the
yearly haul of the proposed Transient Utility will be about 5000
per year.  With such large yields, as discussed by \citep{aof19},
by combining the proposed BTS with higher redshift studies, the
mass of the sum of the masses of neutrinos and uncertainties in the
cosmographic parameters $a$ and $w$ can be constrained.

Incidentally, a standard type Ia at a distance of 400\,Mpc will
peak at 19 mag. Such ``bright"  supernovae can help astronomers
assess the completeness of catalogs of nearby galaxies
\citep{kpm18,fms+19}.\\

\noindent{\bf Transients in the Local Universe.} BTS will favor
bright supernovae such as type Ia supernovae, super-luminous
supernovae (SLSN) and Tidal Disruption Events (TDEs).  A natural
counterbalance to BTS is a volume limited survey, perhaps titled
as ``Transients in the Local Universe" (TILU).  Such a survey will
be sensitive to core-collapse supernovae and other intrinsically
sub-luminous supernovae.  Equally importantly it is the detailed
study of the nearest events (which are necessarily brighter) which
usually provide the best laboratory for understanding the origins.
Such volume limited surveys on a large scale are now being carried
out by ATLAS and ZTF.

The success of TILU will depend directly on the completeness of
catalogs of nearby galaxies \citep{kpm18}.  Fortunately, the era
of massively multiplexed spectrographs has begun. In particular,
Dark Energy Spectroscopic Instrument
(DESI)\footnote{\url{https://www.desi.lbl.gov/}} will, amongst other
surveys, undertake an ambitious spectroscopic survey of bright
galaxies in the Northern sky.

The limiting radius should be large enough to encompass most types
of extragalactic explosions. To this end, I present local rates of
various transients  in Table~\ref{tab:Rates}.  The volume need not
include rare events which are distinctive. Examples include classical
GRBs (distinctive by their high energy spectral and temporal features)
and super-luminous supernovae (distinctive by their long duration
light curves). A radius of 200\,Mpc nicely includes the horizon set
by high energy cosmic rays and by the detection sensitivity of
gravitational wave interferometer for binary neutron star coalescences
\citep{kpm18}.  Again, restricting to transients which go off in
the night sky we expect 1700 transients/year within 200\,Mpc (see
Table~\ref{tab:Rates}).

LSST will play a great role in the TILU survey.  The sensitivity
of LSST means that an explosion in a cataloged nearby galaxy will
be detected at early times.  The primary contamination will be novae
but any transient that is brighter than, say, $-10\,$mag can be
safely assumed to be not a nova. In such cases astronomers can start
undertaking spectroscopic observations with the shallow surveys
providing the full light curves. Next, it appears that some (many)
core-collapse supernovae, across all sub-types, exhibit bumps and
burps before they explode (\citealt{s14} provides a good starting
point for the physics of mass loss in massive stars; see, for
example, \citealt{fr18} for models).  Given this development it is
easy to predict that for all bright supernovae, LSST will provide
a unique insight into this phenomenon.

\subsection{Implementation}

The proposed utility has two elements: an imaging element and
a spectral classification element. 

\subsubsection{Imaging Facilities}
The imaging element requires imaging the entire night sky, at least
once a night,  ideally in several bands, to a limiting magnitude
of 20 mag. This magnitude limit is set by the capability of the
spectroscopic element (described next).

Fortunately, the essential elements for the proposed survey already
exist in the form ATLAS, PS-1 and ZTF. Additionally,  BlackGEM, a
wide-field optical TDA survey based at La Silla \citep{bgn+15}, is
expected to start operating soon. Finally, ATLAS is on course to
adding stations in the Southern hemisphere (J.\ Tonry, pers. comm.).
The (single exposure) sensitivity for these surveys range from 19.5
to 21.5 mag.  Thus, it is a matter for funding agencies or foundations
to entice the operators of these facilities  to participate in a
coordinated public survey.

\subsubsection{Spectroscopic Facilities}
 \label{sec:SpectroscopicFacilities}

For the spectroscopic element there are two components: classification
of bright transients (19\,mag, peak) and fainter transients (say
20\,mag).  For the former, classification observations can be
scheduled to take place close to the peak whereas for the latter,
it would be ideal to observe upon first detection.

From Table~\ref{tab:ClassifyingSpectrographs} we see that the SEDM
has proved itself to be workhorse for 19\,mag events, even though
it is mounted on a 60-inch telescope. The SEDM with its unusual
choice of ultra-low resolution dispersion and robotic acquisition
was built specifically for classifying extragalactic transients.
The spectral classification for BTS can be undertaken with, say,
three 2-m class telescopes (one in the North, one in the South and
one close to the equator and ideally separated in longitude) equipped
with spectrometers tuned for classification. Motivated thus, we are
considering the possibility of an SEDM on the robotic Kitt Peak
84-inch telescope \citep{cdd+19}.

The events in the TILU survey will be, in the mean, fainter than
those of the BTS.  Fortunately, successive generations of spectrographs
have seen an increasing throughput. For instance, consider the Next
Generation Palomar Spectrograph (NGPS), a low resolution spectrograph
that is currently being built for the Palomar 200-inch telescope.
It has a slit-to-detector throughput of 80\% (Figure~\ref{fig:NGPS}).
Two 4-m class telescopes, one in the North and the other in the
South, equipped with  modern low resolution spectrographs, would
be sufficient for this purpose.

In addition to re-prioritizing existing telescope+spectrographs
there is the possibility of new spectroscopic facilities based on
arrays of small telescopes \citep{ebg+19}. In this spirit of novel
approaches, I suggest, based on the experience accrued from the
SEDM that there is a compelling opportunity for considering low
resolution spectrographs optimized solely for SN classification.
Perhaps a funding agency or a foundation could call for a global
competition for innovative solutions to this important and pressing
need. Following the down selection, replicas of the system can be
built and deployed at telescopes, host institutions willing,  around
the world.

 \begin{figure}[ht!] 
 \plotone{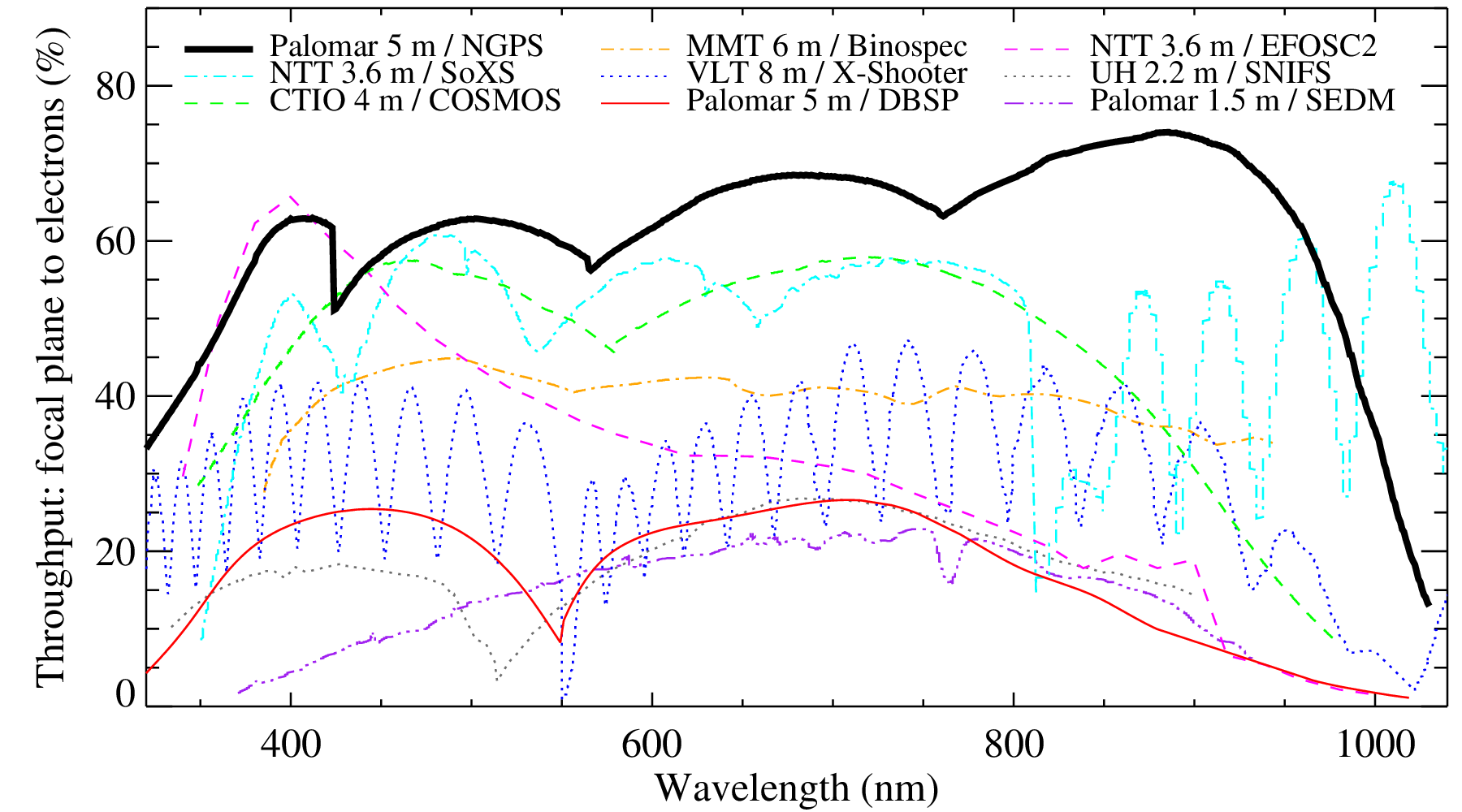} 
 \caption{
Predicted open-slit optical throughput of some of the leading low
to moderate resolution spectrographs.  (SoXS will be commissioned
soon, so its curve is also a prediction.)  The instantaneous
wavelength coverage is significantly narrower than the ranges of
the curves for anything but the lowest resolution modes of COSMOS,
Binospec, and DBSP\@.  SoXS, X-Shooter, and the reciprocal spectrograph
to EFOSC2 (SOFI) extend into the NIR (not shown).  The "sky to
photoelectrons" is displayed for the SEDM.  References: Son of
X-Shooter \citep[SoXS,][M.\ Genoni, pers.\ comm.]{Claudi2018},
COSMOS \citep{Martini2004}, Binospec \citep{Fabricant2019}, X-Shooter
\citep{Vernet2011}, DBSP \citep{Oke82}, EFOSC2, which is part of
PESSTO \citep{Smartt2015}, SNIFS \citep{Lantz2014,Lombardo2017},
and SEDM \citep{bnw+16}.  Figure supplied by E.\ Kirby.  }
  \label{fig:NGPS}
\end{figure}

\section{Conclusions}
 \label{sec:Conclusions}
 
In this report I have focused on the future of  extragalactic
transients, especially those which are  bright enough ($\lesssim
20\,$mag) to allow for routine follow up studies with existing
telescopes.  For fainter transients, much of the study will involve
analysis of multi-band photometric time series from which a highly
selective group will be followed up spectroscopically.

I propose a public ``Optical Transient Utility", based on existing
(and soon to be commissioned) moderate aperture wide-field surveys.
Specifically, a survey(s) which cover the entire night sky and with
sensitivity to readily detected transients with peak magnitude
$\lesssim 19.5$ (which, in practice means, a secure detection at
$\gtrsim 20.5$\,mag). The magnitude limit was chosen so that
spectroscopic classification of most transients can be carried out.
Thus a necessary part of the Utility are two or three 2-m class
telescopes equipped with spectrometers tuned for spectral classification
of supernovae as well as a pair of 4-m class telescopes equipped
with low resolution spectrometers for detailed spectroscopic studies
or early studies of young supernovae (which will be fainter than
19.5\,mag) or both.

The flux limited transient+spectral survey is of considerable value
to (1) continued exploration of the phase space of extragalactic
transients, (2) low redshift cosmology including inferring the
structure of mass on local Baryon Acoustic Oscillation (BAO) scales,
(3) rates of all sub-types of supernovae and tidal disruption events
and (4) accurate measure of the completion fraction of catalogs of
nearby galaxies (few hundred Mpc). In contrast, the volume limited
transient+spectral survey allows for studies of the faint end of
the luminosity function of supernovae and young supernovae.

The primary value of the large projects described above come from
systematic analyses of their vast database.  Fortunately, two
archives have sprung up to meet this demand: the ``Weizmann Interactive
Supernova Data Repository" \citep{yg12} and  the ``Open Supernova
Catalog" \citep{gpk+17}.

\subsection{Critical Role of Astroinformatics}
 \label{sec:Astroinformatics}
 
The rise of TDA surveys was driven by exponential growth in hardware:
format growth of detectors, inexpensive computers for data taking
and analysis and perfection of the detectors. Going forward one of
the avenues for major gains for TDA surveys lie in development and
application of clever algorithms and software methodologies, aka,
``astro-informatics".

Optical TDA surveys are now sufficiently technically mature that
events that they label as transients are genuine astrophysical
events.  In decreasing order they will be supernovae (of all sorts),
eruptive variables (in particular, dwarf novae), flare stars and
asteroids (at turning points). For the moment let us consider the
case of an astronomer interested in supernovae. For this person,
the probability of the candidate not being a supernova is the false
positive probability (FPP).  In contrast, the false negative
probability is the probability that the candidate is a supernova
but the algorithm has classified at not a supernova.

The tolerable level of FPP depends on the circumstances.  Those who are
undertaking a large survey would probably prefer ${\rm FPP \lesssim
0.03}$. On the other hand, someone interested in studying young SNe
may be willing to tolerate ${\rm FPP\approx 0.1}$ so that they will
not miss a ``golden" event.  The key point is that the published
FPP will determine the follow up destiny of the candidate.

\subsection{Clearing Houses}
 \label{sec:ClearingHouses}

To first order we have three types of (apparently and otherwise)
time variable sources: moving objects, variable stars in which I
include eruptive variables, varying AGN and geometric illusions
such as lensing events  and transients.  Fortunately we have two
of the three clearing houses: TNS (\S\ref{sec:Dissemination}) and the
Minor Planet
Center\footnote{\url{https://www.cfa.harvard.edu/content/minor-planet-center-mpc}}
which is ``responsible for the designation of minor planets, comets,
and natural satellites in the solar system. The MPC is also responsible
for the efficient collection, computation, checking, and dissemination
of astrometric observations and orbits for minor planets and comets".

TNS, the clearing house of ATs, is now central to TDA surveys.
Going forward, TNS will play an even more  crucial role. To start
with, as extensively discussed in \S\ref{sec:DiscoveryProcess}, the
``discovery" process will be likely be spread over several groups.
TNS, via appropriate keywords, will have to accommodate the emerging
distributed discovery process. Next, following the discussion in
the previous section TNS should require that groups uploading
candidates to the TNS also  provide FPP. In this spirit, observers/brokers
should be able to update the probabilities as more data are collected.

Given the discussion in the previous paragraph and the analysis in
(\S\ref{sec:Astroinformatics}) ``ratings" agencies are needed to
provide an independent assessment of the quality of candidates
reported by surveys. It would be useful for the appropriate working
group of the IAU (Time Domain Astronomy, perhaps) to discuss this
matter and motivate the community to develop such centers.

At the present time we have no IAU approved clearing house for
variable stars.  In the past, astronomers at the Institute of
Astronomy, Russian Academy of Sciences, maintained the ``General
Catalogue of Variable Stars"\footnote{\url{http://www.sai.msu.su/gcvs/}}
(see \citealt{k16} for a brief history of catalogs of variable
stars). In my view ({\it ibid}), the field of stellar astronomy is
undergoing a second renaissance. If this thesis is accepted then
astronomers would be well served with a major clearing house for
variable stars.  In view of the arguments presented throughout this
paper the variable clearing house should include, in its ambit,
eruptive stars and related phenomena.

\subsection{An Exciting and Endless Frontier}

With little doubt LSST will bring in unanticipated discoveries from
its powerful exploration of the faint events.  LSST is very well
suited  to finding sub-luminous events, both near and far.   Entirely
separately, there are efforts to (1) explore the optical sky on
timescales of seconds and (2) search large swaths of the sky with
clusters of small telescopes for optical counterparts to  GW
sources.\footnote{See for example,
\url{https://evryscope.astro.unc.edu/}\hfil\break
\url{https://goto-observatory.org/}
\url{http://ddoti.astroscu.unam.mx}\hfil\break
\url{https://www.weizmann.ac.il/physics/ofek/research-activities/instrumentation-0}
} Both these developments are being made possible by the availability
of inexpensive but high performance CMOS (Complementary Metal-Oxide
Semiconductor) detectors.  Tomo-e-Gozen is a  massively-mosaicked
CMOS camera with a field-of-view of 20 deg$^2$ and is mounted on
the Kiso 1.05-m Schmidt telescope \citep{sot+18}; see \citet{oss+19}
and \citet{rtm+20} for a taste of the future.

TDA surveys and LSST will drive the need for inexpensive followup
telescopes but equipped with high performance spectrographs (which
are discussed in \S\ref{sec:SpectroscopicFacilities}).  For telescopes
there are two avenues,  both of which are scalable: (1) the use of
arrays of conventional small telescopes (e.g.,  \citealt{ebg+19})
and (2) cheaper moderate-size telescopes based on segmented
architecture and other innovations which lightweight the telescope.
A successful example of the latter is the Kyoto University's Semei
3.8-m
telescope\footnote{\url{https://www.kwasan.kyoto-u.ac.jp/general/facilities/seimei/news20180926_en.html}}
located at Okayama Observatory in Japan.

The TDA revolution is already expanding to other wave bands: NIR,
made possible by drop in detector costs (see \citealt{kaa+19}),
high speed mapping facilities in the radio and the SRG mission in
X-rays.  In my opinion, Ultra-violet (UV) is a ``juicy" frontier
(laden with low hanging fruit). It is a pity that NASA has elected
not to  invest in this rich frontier field.

 \acknowledgements
This paper formed the basis of a talk I gave at the IAU-Kavli meeting
``Transients 2020" (Capetown, February 3--8, 2020).  As usual I am
grateful to librarians and curators who maintain the databases that
modern astronomy relies heavily upon.  For this paper I single out
ADS and TNS.

I am most grateful to Ofer Yaron for providing insight into TNS;
Evan Kirby, PI for NGPS, for discussions on throughputs of various
spectrometers; and Christoffer Fremling for supplying me estimated
annual SN discovery rates. I thank Jesper Sollerman, Anna Ho, Steve
Schulze and Mickael Rigault for careful reading. Finally, I am
grateful to the following for discussion or critical feedback or
both: Ilaria Caiazzo, Richard Dekany, Avishay Gal-Yam, Paul Groot,
Matthew Graham, Thomas Matheson, Jakob Nordin, Eran Ofek, Daniel
Perley, Robert Quimby, Armin Rest, Stephen Smartt and Jason Spyromilio.
SRK thanks the Heising-Simons Foundation for supporting his research
in time domain astronomy.

\bibliography{Bottleneck}{}
\bibliographystyle{aasjournal}

\appendix

\section{Criteria for registering Astronomical Transients with the TNS}
 \label{sec:CriteriaATs}

\subsection{PS-1 \&\ ATLAS}
 \label{sec:PS1ATLAS}
 
Queen's University of Belfast (QUB), under the leadership of S.\
Smartt, is in charge of registering transients arising from PS-1
and ATLAS. The alert schema is described in \citet{ssy+20}.  The
stated goal is to ``register candidate extragalactic explosive
transients as AT." To this end, the following types of events are
rejected: ({\it i}) anything coincident with a  Galactic stellar
source; ({\it ii}) obvious Galactic CVs and novae; ({\it iii})
obvious Galactic M-dwarf flares; ({\it iv}) novae in nearby galaxies
(M31, M33, M81 and others); ({\it v}) known AGN or clear variability
in a galaxy core (as opposed to a high contrast flare, which could
be a TDE). At low latitudes, faint stellar-like candidates are
rejected.

ATLAS is pursuing a key project to follow up all candidates,
regardless of the candidate detection survey, that are in the
vicinity (up to 50\,kpc) of known galaxies within 100\,Mpc. This
survey, has already yield the famous AT2018cow source \citep{pgs+18}
and the faintest SN to date (SN2019gsc; \citealt{ssl+20}).

\subsection{Automatic Learning for the Rapid Classification of
Events (ALeRCE)}
 \label{sec:ALeRCE}

ALeRCE\footnote{\url{http://alerce.science/}} uses a convolutional
neural network  (Carrasco-Davis et al.\ 2020, in prep.) to classify
ZTF public alerts into five classes: asteroid, AGN, variable star,
supernova and bogus. The focus of ALeRCE is to enable fast follow
up of young SNe and as such classification is undertaken even if
there is only one observation.

High-value candidates are automatically displayed at the 
``SNhunter" site\footnote{\url{https://snhunter.alerce.online}}, only
seconds after being ingested from the ZTF  alert stream. After the
end of night in California, three people  (``checkers") visually
inspect the top one hundred SN candidates.  The following events
are registered to TNS: events which are near-extended sources
(putative host galaxies) and which have a PSF-like shape (based on
the ZTF stamps and reference PS-1 images at the location of the
candidate); which are not near known solar system objects; and,
finally, which have not been vetoed by any of the checkers.

NED, Simbad and SDSS are queried to review host galaxy properties
(including redshift, if available). A lower limit for the magnitude
rise rate is computed and reported if the rate exceeds  0.05 mag/day.
Using these tools the group detected  three SN candidates rising
faster than one mag/day:
ZTF19abueupg\footnote{\url{http://dev.alerce.online/object/ZTF19abueupg}},
ZTF19abvdgqo\footnote{\url{http://dev.alerce.online/object/ZTF19abvdgqo}},
ZTF20aaelulu\footnote{\url{http://dev.alerce.online/object/ZTF20aaelulu}}
which were later classified as type II, Ib and Ic SNe, respectively.
Between August 2018 and January 2020 the group has reported 1846
SN candidates with an average rate of 9.1 SNe/day. About two-third
of them have shown a subsequent detection, while one third show
just the first detection due to either being a very distant SNe
near peak, an asteroid near an extended source, or a bogus candidates
mistaken to be real. Over this period, a total 257 of the SN
candidates reported to TNS have been spectroscopically classified
by other groups.

\subsection{Alert Management, Photometry, and Evaluation of Light
Curves (AMPEL)}
 \label{sec:AMPEL}

The AMPEL software platform has been developed to consistently apply
analysis schema to heterogeneous data streams \citep{nbv+19}. One
of the applications has been to perform a selection of likely
extragalactic transients from the ZTF alert stream and share these
with the community through the TNS. The selection methodology was
designed to simultaneously be reproducible, of high quality and
have the capability to submit candidates in real-time to allow for
fast follow-up. These goals were achieved through a fully automated
selection process based on both alert properties and matches with
static astronomical catalogs.  As no human input is needed, the
results carry no unknown selection biases and submission takes place
within minutes of the initial ZTF observations. The selection
quality, including false-positive and completeness rates, were
examined through comparisons with data collected over the first
months of ZTF operations (see \citealt{nbv+19} for a full account).

Currently, two AMPEL TNS “senders” are active: \texttt{ZTF\_AMPEL\_NEW}
only submits candidates with an age less than five days at the time
of submission, thus catering to observers looking for quickly varying
transients.  \texttt{ZTF\_AMPEL\_COMPLETE}has no age criteria and
is instead designed to create a complete, pure candidate stream but
where the average age at submission will naturally be higher. Both
senders operate in parallel and in addition to other TNS senders,
thus allowing users to base follow-up decisions also on which senders
submitted a transient.  The two \texttt{ZTF\_AMPEL} senders have
to date submitted $\sim 8000$ astronomical transients to the TNS,
out of which $\sim 1900$ were later classified as SNe. ``The vast
majority of the unclassified transients appear to be real but fainter
transients."

\subsection {Bright Transient Survey (ZTF)}
 \label{sec:ZTF-BTS}
 
The Bright Transient Survey (BTS) is the largest SN survey undertaken
by ZTF. This survey is described in \citet{fms+19}.  After the first
year of the survey, the original simple rubric was replaced with a
more complex series of cuts to decrease the reliance on human
scanning and judgment (which by mid-2019 after reference imaging
was completed, was requiring scanners to pick out ~5 SNe out of
400+ candidates per night).  The current filter, like the previous
one, requires that sources be above the reference level and above
19.0 mag, that they be detected in two or more epochs separated in
time by at least 30 minutes, and that they be outside the Galactic
plane ($\vert b\vert>7^\circ$).  Additionally, (1) sources with a
minor planet match are explicitly rejected.  (2) Sources are rejected
if they have a ML real-bogus score less than 0.2.
This threshold is increased to 0.35 within $<1^{\prime\prime}$ of
$<17.0\,$mag {\it Gaia}-catalogued objects and to 0.45 within
$<1.5^{\prime\prime}$ of $<15.5$\,mag {\it Gaia} objects.  (3) Sources
must also not have a "deep" ML score less than 0.1.
(4) Sources are required to not be coincident within $<2^{\prime\prime}$
of a high-probability star (ML ``stellarity" score
$>0.76$), or within $<0.5^{\prime\prime}$ of a bright ($<17\,$mag)
object for which star/galaxy classification failed, or within
$<1^{\prime\prime}$ of an extremely red stellar object ($r-z >
3\,$mag with stellarity $>0.2$, a probable M-dwarf).  They must
also not be within $1.5^{\prime\prime}$ of a $<15\,$mag {\it Gaia} object
or within $1.0^{\prime\prime}$ of a $<16.5\,$mag {\it Gaia} object, but
only if the source is not new ($<15\,$days and $<30\,$ days,
respectively).  (5) Sources even moderately close to very bright
stars are rejected due to the frequency of artifacts in these
regions.  The event cannot be within $20^{\prime\prime}$ of a
$r<12\,$mag probable star (stellarity $>0.49$) or a $r<14.5\,$mag
definite star (stellarity $>0.8$), within 5$^{\prime\prime}$ of a
$r<15\,$mag probable star, or within $1.1^{\prime\prime}$ of a
$r<16.5\,$mag probable star.  (Similar cuts are also applied with
reference to $i$-band and $z$-band mags.)

Sources with a long history (time from first detection) are scrutinized
more carefully, since the vast majority of such sources are AGNs
or variable stars.  If within $3^{\prime\prime}$ of a $<16\,$mag
catalogued source and older than $>90\,$days, the source is rejected.
``Old" sources ($>90\,$days) are also rejected if $<0.4^{\prime\prime}$
from a $<19.5\,$mag catalogued source, $<0.8^{\prime\prime}$ from
a $<17.5\,$mag catalogued source, or $<1.2^{\prime\prime}$ from a
$<15.5\,$mag catalogued source.  However, this rejection does not
apply if the source is at a maximum in its light curve or if it has
not been recorded at $<18.5\,$mag more than twice in the past 30
days, to avoid missing SNe near AGNs or galaxy nuclei.   Sources
very closely coincident with {\it Gaia}-catalogued objects are also removed
if  ``old" (specifically: $<0.35^{\prime\prime}$ from a $<17\,$mag
object if $>30\,$days, $<0.20^{\prime\prime}$ from a $<18\,$mag
object if $>90\,$days, and $<0.35^{\prime\prime}$ from a $<19\,$mag
object if $>300\,$days and detected fainter than 18.5\,mag).

Approximately 50 new sources pass these criteria during a clear
night, of which about 5 are typically SNe and the rest are a mix
of CV's, AGNs variable stars, and a handful of artifacts which still
pass through the above cuts for various reasons.  (These numbers
do not include sources that have already been saved during previous
nights). Human scanners, aided by the GROWTH Marshal \citep{kcb+19},
using contextual information (typically whether the source is located
within a galaxy and not coincident with a central point source, and
whether its light curve is varying smoothly or erratically), reject
the interlopers.  Only a few minutes are required to scan and
identify these sources on a particular night.

\subsection{Gaia}
 \label{sec:Gaia}
 
{\it Gaia Science Alerts}
system\footnote{http://gsaweb.ast.cam.ac.uk/alerts} \citep{Wyrzykowski2012,
Hodgkin2013} is a part of the European Space Agency's {\it Gaia}
space mission \citep{Gaia2016} and is operated by the Institute of
Astronomy of the University of Cambridge, UK, with the support of
researchers from the University of Warsaw, Poland, SRON, NL and
Konkoly, Hungary.  {\it Gaia}, which has been operational since
2014, covers about 1000 deg$^2$ over 24 hours and its data is
transmitted to the ground on average every day. The typical delay
between an observation and an alert is about 24--48 hours and typical
cadence is 106 mins (due to the two mirror geometry of the instrument)
and 30 days (due to scanning pattern). Alerts are defined as
significant changes as measured by {\it Gaia} photometers (G-band;
broad band optical). In the first two years of operation, {\it Gaia}
Alerts delivered on average 1 alert per day, while now the rate is
more than 10 transients/day.  The detection pipeline is unbiased
and is sensitive to both new sources (e.g., supernovae, cataclysmic
variables) and old sources (e.g., micro-lensing events, AGN flares).

The pipeline developed over years of the operation is capable of
removing artifacts due to e.g., diffraction spikes from bright
stars, or overlapping windows for stars in crowded regions. Note
that {\it Gaia} mission processes its images on-board and what is
delivered to the ground are positions and fluxes of detected objects
brighter than about 20.5 mag.  {\it Gaia} Alerts positional accuracy
is about $0.1^{\prime\prime}$ but that improves to tens of
micro-arcseconds for reprocessed data published in subsequent data
releases \citep{GaiaDR2}. A unique capability of {\it Gaia} is its
low-dispersion (R$\sim$70--100) spectrophotometric measurements
obtained for every individual observation with its Blue and Red
Photometers (BP-RP), opening new possibilities for detection and
classification of transients based on spectra, not just fluxes
\citep{Blagorodnova2014}.

\subsection{\it Optical Gravitational Lensing Experiment (OGLE)}
 \label{sec:OGLE}

OGLE has been operational since  1992. However, only since 2012 it
started to deliver regular transient detections  \citep{Wyrzykowski2014}.
OGLE monitors about 700 deg$^2$ of the sky, located around and
between Magellanic Clouds, with a cadence varying from days to
weeks. The OGLE Transient Detection System
\footnote{http://ogle.astrouw.edu.pl/ogle4/transients/} uses
difference imaging method and can observe transients as faint as
21.5 mag in I-band. Its long-term imaging history allows for detection
of slowly evolving extragalactic transients.

I thank the following for contributing text to the sections above:
S.\ Smartt (\S\ref{sec:PS1ATLAS}), F.\ Bur\'on (\S\ref{sec:ALeRCE}),
J.\ Nordin (\S\ref{sec:AMPEL}), D.\ Perley (\S\ref{sec:ZTF-BTS})
and L.\ Wyrzykowski (\S\ref{sec:Gaia} \&\ \S\ref{sec:OGLE}).


\begin{deluxetable}{lr}[hbt]
\tablecaption{Histogram of Host Galaxy Catalogs}
\tablewidth{0pt}
\tablehead{
\colhead{Galaxy Catalog} & \colhead{$n_g$}
}
\startdata
\input{GalaxyCatalog.txt}
\enddata
\tablecomments{Column 1 is the entry for galaxy catalogs and 
$n_g$ (column 2) is the number of galaxies in that catalog. The
histogram has been cutoff at $n_g<10$.}
 \label{tab:HostGalaxies}
\end{deluxetable}

\section{Additional Analyses}

\subsection{Host Galaxies}
 \label{sec:HostNames}
 
Some reporting astronomers provide host galaxy information (``Host
Name", field 7).  For the 2019 dataset, it appears that 1819 events,
composed of 409 SNe and 1410 ATs, have this field populated.The
histogram of the host galaxy catalogs is given in
Table~\ref{tab:HostGalaxies}.

\subsection{Bright Transient Candidates}
 \label{sec:BrightTransientCandidates}

%

In this section I investigate why so many of the bright transients
were not classified.  I carried out an analysis of ATs for the
sample of ATs with discovery magnitude $<17.5$. There are 1608 such
events, of which only 225 are classified. The latter set is composed
of 161 SNe, 45 CVs and a smattering of ``Galaxy" (which means that
the classification spectrum was dominated by galaxy light), ``variable
stars", ``novae" and ``AGN". For the remaining 1383 ATs the counts
from the reporting agents  can be found in
Table~\ref{tab:BrightUnclassified}. Given {\it Gaia} Alerts philosophy
of reporting all strong variables (\S\ref{sec:Gaia}) it is not
surprising that the leading contributor of bright unclassified ATs
is {\it Gaia}. MASTER also has a similar philosophy as {\it Gaia},
namely, to report strong variables, eruptive variables and transients
to TNS.


There are five different ZTF ``bots" or channels which upload
candidates to TNS and the accounting is as follows:
\texttt{ZTF\_AMPEL\_NEW} (160/16), \texttt{ZTF\_AMPEL\_COMPLETE}
(126/2), \texttt{ZTF\_Bot1} (67/6), \texttt{ZTF\_CLU\_PUBLIC} (1/0)
and \texttt{ZTF\_TESS} (2/0). Here, for each enclosed value the
first value is the number of alerts loaded and the second is the
number that is classified. Of the five, only \texttt{ZTF\_CLU\_PUBLIC}
is human vetted.  Nonetheless, given that ZTF and ATLAS are focused
on genuine transients it is rather surprising to find large
contributions to this Table  from ZTF and ATLAS.  ASAS-SN reports
human-vetted events and even so the number of classifications is
91 whereas the number of reported events is 133.


\begin{deluxetable}{lr}[hbt]
\tablecaption{Histogram of Bright ($<17.5\,$mag) Unclassified ATs}
\tablewidth{0pt}
\tablehead{
\colhead{Reporting Group} & \colhead{$n_r$}
}
\startdata
\input{BrightEventsTable.txt}
\enddata
\tablecomments{Column 2 is the number of unclassified bright
($<17.5\,$mag) ATs reported by group (Column 1).  ATs with field 9
(reporting group) are assigned ``uno". Most of them are detections
reported by amateur  astronomers.  The histogram has been cutoff
at $n_r<5$.}
 \label{tab:BrightUnclassified}
\end{deluxetable}

\section{SEDM: An Ultra-low Resolution Spectrograph tuned for SN
classification} 
 \label{sec:SEDM}

Traditionally, spectral classification has been done with ``Low
resolution" optical spectrographs such as the famed Double Beam
Spectrograph (DBSP) on the 200-inch telescope or the Double Imaging
Spectrograph on the 3.5-m telescope of the Apache Point Observatory.
The spectral resolution, $R=\lambda/\Delta\lambda$, of such
spectrographs is a few thousand; here $\Delta\lambda$ is the wavelength 
spread of the slit spread function.  With this resolution it is easy
to distinguish active stars (flare stars, RS CVn etc.), cataclysmic
variables, supernovae, tidal disruption events and burping AGNs.
However, when detectors have read noise (as is still the case with
current generation of CCDs) then spectral resolution comes with a
cost.  The larger the spectral resolution, longer is the minimum
exposure time or equivalently one needs a larger telescope (for the
same exposure time).

Supernovae  dominate the demographics of extragalactic transients.
The expansion speeds of supernovae $v\gtrsim 3,000$\,km\,s$^{-1}$.
Thus a spectral resolution of $R\approx c/v$ or one hundred is
adequate to classify supernovae and tidal disruption events. This
resolution is certainly inadequate for spectral classification of
stars.

Motivated thus we constructed an {\it ultra-low} resolution
spectrograph with $R\approx 100$. SEDM is  the sole instrument on
the Palomar 60-inch telescope (henceforth, ``SEDMv1"). In order to
cut down on acquisition time we opted for an integral field unit
(IFU)  with a wide entrance field-of-view of $30^{\prime\prime}$
on the side.  Given the low resolution and the expected high target
throughput we called it as the ``Spectral Energy Distribution
Machine" (SEDM). The hardware and initial performance details can
be found in \citet{bnw+16}.  Data reduction of IFU data is very
tricky and subtle. The details of the data reduction pipeline can
be found in \citet{rnb+19}.

\end{document}

%% file: AT_statistics.txt
ZTF & 6922\\
PS-1 & 5324\\
Gaia & 2929\\
ATLAS & 2211\\
ASAS-SN & 256\\
DECam & 190\\
MASTER & 122\\
OGLE & 109\\

%% file: TelescopeSpectrograph.txt
P60 - SEDM                &  810  & 40.3 \cr 
ESO-NTT - EFOSC2-NTT      &  294  & 14.6 \cr 
LT - SPRAT                &  215  & 10.7 \cr 
P200 - DBSP               &  154  & 7.7 \cr 
anon                      &   95  & 4.7 \cr 
UH88 - SNIFS              &   94  & 4.7 \cr 
Lick-3m - KAST            &   63  & 3.1 \cr 
APO-3.5m - DIS            &   42  & 2.1 \cr 
Ekar - AFOSC              &   41  & 2.0 \cr 
FTN - FLOYDS-N            &   32  & 1.6 \cr 
SOAR - Goodman            &   30  & 1.5 \cr 
LCO-duPont - WFCCD        &   23  & 1.1 \cr 
Lijiang-2.4m - YFOSC      &   22  & 1.1 \cr 
FTS - FLOYDS-S            &   18  & 0.9 \cr 
Magellan-Baade - IMACS    &   16  & 0.8 \cr 
Magellan-Clay - LDSS-3    &   10  & 0.5 \cr 

%% file: GalaxyCatalog.txt
SDSS & 576\\
WISE & 353\\
2MASS & 339\\
OGLE & 96\\
GALEX & 58\\
NGC & 53\\
LEDA & 47\\
UGC & 45\\
CGC & 34\\
ESO & 29\\
MCG & 19\\
M31 & 18\\
PGC & 17\\
IC & 17\\
KUG & 15\\

%% file: BrightEventsTable.txt
  GaiaAlerts & 798 \\
         ZTF & 331 \\
       ATLAS & 75 \\
      MASTER & 70 \\
     ASAS-SN & 46 \\
         uno & 8 \\
 Pan-STARRS1 & 7 \\
       BraTS & 7 \\
      ALeRCE & 7 \\
        XOSS & 6 \\
       GSNST & 6 \\
     Gattini & 5 \\